\documentclass[12pt]{article}
\usepackage{amsmath} 

\title{Emergent diffeomorphism invariance in toy models}
\author{Hrvoje Nikoli\'c \\
Theoretical Physics Division, Rudjer Bo\v{s}kovi\'{c} Institute, \\
P.O.B. 180, HR-10002 Zagreb, Croatia \\
{\normalsize e-mail: hnikolic@irb.hr} \\
\makebox[1in]{} \\
}
\date{\today}
\begin{document}
\maketitle
\begin{abstract}
Conceptual difficulties in semiclassical and quantum gravity arise from diffeomorphism invariance 
of classical general relativity. With a motivation to shed some light on these difficulties, we study a class of 
toy models 
for which one-dimensional diffeomorphism invariance, 
namely time-reparametrization invariance, emerges at the classical level from energy conservation.
An attempt to quantize the models while taking the invariance seriously leads to toy versions of the 
problem of time in quantum gravity, of the cosmological constant problem, and of the black hole firewall problem. 
Nevertheless, all these problems are easily resolved by taking into account that 
the invariance emerges only at the classical level, while the fundamental theory that needs to be quantized is not 
diffeomorphism invariant.      
\end{abstract}
\vspace*{0.5cm}


\section{Introduction}

Classical general relativity \cite{mtw,wald,carroll} is one of the most elegant theories in physics. 
Its most distinguished feature is diffeomorphism invariance, or invariance under active 
general transformations of spacetime coordinates, which implies that spacetime metric is a dynamical quantity.
But this elegance is a blessing and a curse. It's a blessing in classical physics, but a curse in quantum physics 
because we still do not fully understand how to quantize gravity \cite{kiefer,rovelli,becker}, that is,  
how to implement diffeomorphism invariance at the quantum level. The problems appear not only in fully 
quantum gravity, but also in the semiclassical approximation \cite{bd,mukhanov} where only matter is quantized
while gravity is treated classically. The problems that appear are not only technical, but also conceptual.    
The three conceptual problems that stand out are 
the problem of time in quantum gravity \cite{kuchar,isham,padm_timeqg,anderson},
the cosmological constant problem \cite{weinberg,nobb,sahni,carroll_cc,padm_cc}, 
and the black hole information paradox 
\cite{gid,har,pres,pag,gid2,str,math1,math2,hoss,dundar,harlow,polchinski,chakra,marolf,fabbri}.

One possibility that potentially could help to resolve these conceptual problems is the idea that 
general relativity and its diffeomorphism invariance is emergent, rather than fundamental,
while the underlying more fundamental theory rests on entirely different principles.
This idea can be realized in condensed-matter inspired theories such as induced gravity \cite{sakharov},
as well as in string theory \cite{becker}. However, there is no any direct experimental evidence
for such a more fundamental theory. Moreover, promising theoretical candidates such as string theory
are still poorly understood in their most fundamental terms. Consequently, it is very difficult to study the idea
of emergent diffeomorphism invariance
in realistic models. In this paper, therefore, we study this idea in toy models, similar to the toy models 
in \cite{padm_timeqg,kuchar,nik_donder-weyl} studied before in the context of the problem of time in quantum gravity.
In these models, the 4-dimensional spacetime diffeomorphism invariance of general relativity 
is replaced with a 1-dimensional diffeomorphism invariance realized as time-reparametrization invariance. 
Even though such models cannot solve the problems of realistic 4-dimensional systems with gravity,
it is hoped that such simple models can at least serve as a conceptual inspiration for dealing 
with more difficult realistic theories.

The paper is organized as follows. In Sec.~\ref{SECmodel_and_emerg} we first introduce a class of 
toy models 
without diffeomorphism invariance and then explain how 1-dimensional diffeomorphism invariance emerges 
from conservation of energy, namely, as a way to implement the constraint that the classical system has definite 
energy. 
In Sec.~\ref{SECtimeqg} we explain how the 1-dimensional diffeomorphism invariance leads to a 
toy version 
of the problem of time in quantum gravity, and how the problem resolves when one recalls that the diffeomorphism invariance
is not fundamental.  
Similarly, in Sec.~\ref{SECcc} we explain how the 1-dimensional diffeomorphism invariance leads to a 
toy version 
of the cosmological constant problem, 
and how the problem resolves when one recalls that the diffeomorphism invariance is not fundamental.
Likewise, in Sec.~\ref{SECbh} we find a solution of the constraint that in some aspects resembles the behavior in a black hole 
exterior, explain how the diffeomorphism invariance can be used to extend the solution to a region 
resembling the behavior in a black hole interior, and point out that the interior is actually unphysical 
because the diffeomorphism invariance is not fundamental. The non-existence 
of the interior can be understood as a toy version of the black hole firewall \cite{AMPS,apologia}, 
which plays a key role in some approaches to solving the black hole information paradox.
In Sec.~\ref{SEC4dim} we briefly speculate how these toy models could perhaps be generalized 
to real 4-dimensional diffeomorphism invariance. 
Finally, in Sec.~\ref{SECconcl} we present a qualitative discussion of our results.

\section{The model and emergent diffeomorphism invariance}
\label{SECmodel_and_emerg}

\subsection{The model}
\label{SECmodel}

We study a system with $N$ dynamical degrees of freedom described by the collective configuration variable
$q(t)=\{ q_1(t),\ldots,q_N(t) \}$, the dynamics of which is described by the action
\begin{equation}\label{A}
 A=\int dt\, L(q,\dot{q}) , 
\end{equation}
where the dot denotes the derivative with respect to time $t$ and 
\begin{equation}\label{L}
  L(q,\dot{q})=\sum_{a=1}^N \frac{m_a\dot{q}_a^2}{2}-V(q) .
\end{equation}
The canonical momenta are well defined
\begin{equation}
 p_a=\frac{\partial L}{\partial \dot{q}_a} = m_a \dot{q}_a ,
\end{equation}
so the Hamiltonian is 
\begin{equation}
  H(q,p)= \sum_{a=1}^N p_a\dot{q}_a - L = \sum_{a=1}^N \frac{p_a^2}{2m_a}+V(q) 
\end{equation}
and can be interpreted as the energy of the system.
The system can be treated either classically of quantum mechanically, in a straightforward manner.
In particular, quantization can be performed via canonical quantization and dynamics 
can be described by the Schr\"odinger equation
\begin{equation}\label{sch}
 H|\psi(t)\rangle = i\hbar \partial_t |\psi(t)\rangle 
\end{equation}
as usual, where $H$ is the operator.
Since the action does not have any {\it a priori} gauge or diffeomorphism invariance, 
the quantization is straightforward.

\subsection{Emergent diffeomorphism invariance}
\label{SECemerg}

Since the Hamiltonian $H$ does not have an explicit time dependence, it is conserved. In classical physics, this means 
that $H$ has some definite constant value $E$ of energy, so we can write it as $H(q,p)=E$, or
\begin{equation}\label{constr1}
 {\cal H}(q,p)=0 ,
\end{equation}
where 
\begin{equation}
 {\cal H}(q,p) \equiv H(q,p) - E .  
\end{equation}
In the configuration space, the fact that the Hamiltonian has the value $E$ can be written as 
\begin{equation}\label{constr2}
 \sum_{a=1}^N \frac{m_a\dot{q}_a^2}{2} + V(q) -E =0 .
\end{equation}
If we imagine that (\ref{L}) describes a whole Universe, then $E$ is the energy of that Universe.
The inhabitants of this Universe observe only one value of $E$, but the theory cannot say which one. 
For the inhabitants of this Universe, the constant $E$ is a fundamental constant the value of which 
can be determined from experiments. 

Since $E$ appears as a fundamental constant, it seems natural to incorporate the value of this constant into an 
effective action. One possibility is to incorporate the constraint (\ref{constr2}) into the action by adding 
the Lagrange multiplier term $\lambda \left[ \sum_{a}m_a\dot{q}_a^2/2 + V(q) -E \right]$. 
However, there is a much more interesting way to incorporate the constraint (\ref{constr2}) into the action.
We do that not by introducing a Lagrange multiplier $\lambda$, but by introducing a new  
configuration variable $g(t)>0$ and replacing the action (\ref{A}) with 
\begin{equation}\label{A2}
 \tilde{A}=\int dt\sqrt{g} \left[ \sum_{a=1}^N \frac{m_a\dot{q}_a^2}{2g}-V(q) +E \right] .   
\end{equation}
Since this action does not depend on time derivatives of $g(t)$, the $g(t)$ is not a dynamical variable 
and the equation of motion for this variable is a constraint equation. More precisely, the equation of motion 
$\delta\tilde{A}/\delta g=0$ gives
\begin{equation}\label{constr3}
 -\frac{1}{2\sqrt{g}} \left[ \sum_{a=1}^N \frac{m_a\dot{q}_a^2}{2g} + V(q) -E \right]=0 ,
\end{equation}
which reduces to the constraint (\ref{constr2}) if $g=1$. But what is the rational for taking $g=1$?
The answer is that the action (\ref{A2}) has the property of {\em diffeomorphism invariance} which allows us to 
choose for $g(t)$ any positive function we want, so $g(t)=1$ is nothing but a convenient choice of ``gauge''. 
Since this diffeomorphism invariance is crucial, let us explain it in more detail.

The $g$ in (\ref{A2}) appears in two terms, which are proportional to
\begin{equation}
 dt\sqrt{g} , \;\;\;\; \frac{\dot{q}_a^2}{g}=\frac{dq_a^2}{g\,dt^2} .
\end{equation}
Thus $g$ appears either in the combination $\sqrt{g}dt=\sqrt{g\,dt^2}$ or $g\,dt^2=(\sqrt{g}dt)^2$. 
This implies that the action is invariant under arbitrary transformations that keep
\begin{equation}\label{dtau2}
 d\tau^2\equiv g(t)dt^2
\end{equation}
invariant. The $d\tau^2$ is very much analogous to the spacetime line element $ds^2 = g_{\mu\nu}(x)dx^{\mu}dx^{\nu}$
in general relativity, so we see that $g$ in (\ref{dtau2}) corresponds to $g_{00}$ in general relativity. 
Likewise, $1/g$ corresponds to $g^{00}$. Just like general relativity is invariant under arbitrary 4-dimensional 
spacetime diffeomorphisms $x^{\mu}\to x'^{\mu}=f^{\mu}(x)$ which keep $ds^2 = g_{\mu\nu}(x)dx^{\mu}dx^{\nu}$
invariant, the action (\ref{A2}) is invariant under arbitrary 1-dimensional time diffeomorphisms
\begin{equation}\label{t'} 
 t\to t'=f(t)
\end{equation} 
which keep (\ref{dtau2}) invariant. The invariance $g\,dt^2=g'dt'^2$ implies that $g$ transforms as 
\begin{equation}
 g\to g'=\left( \frac{dt}{dt'} \right)^2 g.              
\end{equation}
This 1-dimensional diffeomorphism invariance is also known in literature under the name 
{\em time-reparametrization invariance} \cite{rovelli,anderson,isham}.   
 
To summarize, we have started from the action (\ref{A}) without diffeomorphism invariance and,
from the fact that energy has some constant value $E$ in classical mechanics, 
derived the corresponding action (\ref{A2}) {\em with} 1-dimensional diffeomorphism invariance. 
In this way, the 1-dimensional diffeomorphism invariance is {\em emergent from classical energy conservation}.

\subsection{The constraint in the canonical form}

Now we want to develop some formal tools that will be used in further sections.
The action (\ref{A2}) can also be written as 
\begin{equation}\label{A3}
 \tilde{A}=\int dt\, \tilde{L}(q,\dot{q},g) = \int dt\sqrt{g}\, {\cal L}(q,\dot{q},g),   
\end{equation}
where
\begin{eqnarray}\label{A3defs} 
& {\cal L}(q,\dot{q},g) = \displaystyle\sum_{a=1}^N \frac{m_a\dot{q}_a^2}{2g}-V(q) +E , &
\nonumber \\
& \tilde{L}(q,\dot{q},g) = \sqrt{g}{\cal L}(q,\dot{q},g) . &
\end{eqnarray}
The corresponding canonical momenta are 
\begin{equation}\label{momenta}
 \tilde{p}_a=\frac{\partial\tilde{L}}{\partial \dot{q}_a}=\frac{m_a\dot{q}_a}{\sqrt{g}} , \;\;\;
p_g=\frac{\partial\tilde{L}}{\partial \dot{g}}= 0 ,
\end{equation}
so the Hamiltonian is
\begin{equation}
 \tilde{H}(q,\tilde{p},g)=\sum_{a=1}^N \tilde{p}_a\dot{q}_a - \tilde{L}= \sqrt{g}\, {\cal H}(q,\tilde{p}) ,
\end{equation}
where
\begin{equation}
 {\cal H}(q,\tilde{p}) = \sum_{a=1}^N \frac{\tilde{p}_a^2}{2m_a}+V(q)-E .
\end{equation}
The canonical equation of motion for $p_g$ is
\begin{equation}\label{dotpg}
 \dot{p}_g=-\frac{\partial\tilde{H}}{\partial g}=-\frac{1}{2\sqrt{g}}{\cal H} .
\end{equation}
However, in (\ref{momenta}) we have seen that $p_g=0$, which implies $\dot{p}_g=0$, so 
(\ref{dotpg}) implies
\begin{equation}\label{constr3can}
 -\frac{1}{2\sqrt{g}}{\cal H}=0 ,
\end{equation}
which is identical to the constraint (\ref{constr3}). Thus, since $g>0$,
we see that the constraint (\ref{constr3}), or (\ref{constr3can}), 
can also be written as the Hamiltonian constraint
\begin{equation}\label{constr4}
 {\cal H}(q,\tilde{p})=0 ,
\end{equation}
or equivalently
\begin{equation}\label{constr5}
 \tilde{H}(q,\tilde{p},g)=0 .
\end{equation}
In the gauge $g=1$, this reduces to the constraint (\ref{constr1}). 

\section{The problem of time in quantum gravity}
\label{SECtimeqg}

Seduced by the beauty and elegance of the action with 1-dimensional diffeomorphism invariance, one may be tempted 
to quantize it. The problem is, how to implement the Hamiltonian constraint (\ref{constr4}) in the quantum theory?
The most natural approach is to implement it as the constraint on physical states
\begin{equation}
{\cal H}(q,\tilde{p})|\psi\rangle = 0 , 
\end{equation}
where ${\cal H}(q,\tilde{p})$ is the quantum operator obtained via standard canonical quantization. 
This constraint implies also
\begin{equation}\label{sch0}
 \tilde{H}(q,\tilde{p},g)|\psi\rangle = 0 ,
\end{equation}
which is the quantum version of (\ref{constr5}).
However, the time evolution of the state should be described by the corresponding Schr\"odinger equation
\begin{equation}
 \tilde{H}(q,\tilde{p},g)|\psi(t)\rangle = i\hbar\partial_t |\psi(t)\rangle ,
\end{equation}
so compatibility with (\ref{sch0}) implies
\begin{equation}
 \partial_t |\psi(t)\rangle = 0 .
\end{equation}
Hence the state does not depend on time. But we know that the real world, 
or even the toy world described by the toy model in Sec.~\ref{SECmodel}, depends on time.
Where does the dependence on time come from, if the quantum state $|\psi(t)\rangle$ does not depend 
on time? This is the toy version of the problem of time in quantum gravity \cite{kuchar,isham,padm_timeqg,anderson}.

Within our model, it is not difficult to understand where the problem comes from 
and how it should be resolved. In general, whenever a quantum system has a well defined energy $E$,
its wave function has trivial time dependence proportional to $e^{-iEt/\hbar}$, 
which is just a time-dependent phase without any physical consequences. To have a genuine time-dependent 
state in quantum mechanics, the state must {\em not} have a well defined energy. Instead, 
the state must be in a superposition of two or more {\em different} energies. 

So what is wrong with (\ref{sch0})? This quantum constraint originates from the classical action 
(\ref{A2}) in which the energy $E$ is {\em fixed}. In fact, the whole diffeomorphism invariance of 
(\ref{A2}) emerged from a desire to implement the classical value $E$ of energy into the action.
There is nothing wrong with it in classical physics, where energy indeed has a well defined value.
However, requiring that the quantum system should also have a definite value of energy is wrong,
because the energy of a quantum system is, in general, uncertain. In other words, it is wrong to quantize 
the diffeomorphism invariant effective action (\ref{A2}). What needs to be quantized is the original 
action (\ref{A}), which is not diffeomorphism invariant and leads to the proper Schr\"odinger equation (\ref{sch})
without the problem of time. The emergent diffeomorphism invariance is only valid at the classical level,
where energy is well defined. At the quantum level, where energy is uncertain, there is no diffeomorphism invariance.      

To conclude, the problem of time in the toy version of quantum gravity originates from taking the 
diffeomorphism invariance too seriously. When one takes into account that this invariance is only emergent 
at the classical level, while fundamental quantum theory does not have this invariance, the problem 
of time disappears in an obvious way.  

\section{The cosmological constant problem}
\label{SECcc}

Among the $N$ degrees of freedom, let us suppose that $N_{\rm heavy}$ of them are ``heavy'' and the rest 
$N_{\rm light}=N-N_{\rm heavy}$ are ``light''. We call them ``heavy'' and ``light'' degrees 
because we assume that one can use a semiclassical approximation in which the $N_{\rm heavy}$ degrees
are treated classically, while the rest $N_{\rm light}$ of them are quantized. For simplicity, 
we also assume that $V(q)$ can be split as 
\begin{equation}
 V(q)=V_{\rm heavy}(q_{\rm heavy}) + V_{\rm light}(q_{\rm light}) ,
\end{equation}
where $q_{\rm heavy}=\{ q_b\,|\, b=1,\ldots,  N_{\rm heavy} \}$ are heavy degrees, and 
$q_{\rm light}=\{ q_a\,|\, a=1,\ldots,  N_{\rm light}\}$ are light degrees. 
Thus the classical constraint (\ref{constr3}) can be written as
\begin{equation}\label{einst1}  
 -\sum_{b=1}^{N_{\rm heavy}}\frac{m_b\dot{q}_b^2}{2g} - V_{\rm heavy}(q_{\rm heavy})  
= \sum_{a=1}^{N_{\rm light}} \frac{m_a\dot{q}_a^2}{2g} + V_{\rm light}(q_{\rm light}) -E ,
\end{equation}
or more concisely
\begin{equation}\label{einst2}
 -{\cal H}_{\rm heavy} = {\cal H}_{\rm light} - E,
\end{equation}
with a self-explaining notation. This is a classical equation, but as we said, the idea is to treat it semi-classically,
so that the light degrees are quantized while the heavy degrees are left classical. Thus one replaces 
(\ref{einst2}) with a semiclassical equation
\begin{equation}\label{einst3}
 -{\cal H}_{\rm heavy} = \langle\psi| {\cal H}_{\rm light} |\psi\rangle - E,
\end{equation}
where $\langle\psi| {\cal H}_{\rm light} |\psi\rangle$ is the mean value of the operator ${\cal H}_{\rm light}$ 
in the quantum state $|\psi\rangle$. 

Next suppose that $V_{\rm light}(q_{\rm light})$ is the potential of $N_{\rm light}$ harmonic oscillators
\begin{equation}
 V_{\rm light}(q_{\rm light})=\sum_{a=1}^{N_{\rm light}} \frac{k_a q_a^2}{2} .
\end{equation}
Then the operator ${\cal H}_{\rm light}$ can be written in the usual quantum harmonic oscillator form
\begin{equation}
 {\cal H}_{\rm light}=\sum_{a=1}^{N_{\rm light}} \hbar\omega_a \left( A^{\dagger}_a A_a +\frac{1}{2} \right) , 
\end{equation}
where $\omega_a=\sqrt{k_a/m_a}$, while $A^{\dagger}_a$ and $A_a$ are the raising and lowering operators, 
respectively. In particular, in the quantum ground state defined by $A_a|0\rangle=0$ we have
\begin{equation}\label{Hvac}
\langle 0| {\cal H}_{\rm light} |0\rangle=\sum_{a=1}^{N_{\rm light}} \frac{\hbar\omega_a}{2} ,
\end{equation}
so the semiclassical equation (\ref{einst3}) becomes
\begin{equation}\label{einst4}
 -{\cal H}_{\rm heavy} = \sum_{a=1}^{N_{\rm light}} \frac{\hbar\omega_a}{2} - E.
\end{equation}
By contrast, the ground state energy of the classical harmonic oscillator is zero, so the classical 
version of (\ref{einst4}) is 
\begin{equation}\label{einst5}
 -{\cal H}_{\rm heavy} = - E.
\end{equation}
But $N_{\rm light}$ is supposed to be very large, after all this is the number of light degrees in the 
{\em whole} toy Universe. Thus, there is a large discrepancy between the classical equation (\ref{einst5})
and the semiclassical equation (\ref{einst4}). The semiclassical equation (\ref{einst4}) can also be written as 
\begin{equation}\label{einst6}
 -{\cal H}_{\rm heavy} = - E_{\rm eff}, 
\end{equation}
where 
\begin{equation}\label{Eeff0}
 -E_{\rm eff}=-E+\sum_{a=1}^{N_{\rm light}} \frac{\hbar\omega_a}{2} .
\end{equation}
The effective energy $E_{\rm eff}$ contains a very large contribution from the quantum zero-point energy.

Finally, suppose that the inhabitants of the toy Universe measure $E_{\rm eff}$ and find a value 
\begin{equation}\label{Eeff}
 -E_{\rm eff} \ll \sum_{a=1}^{N_{\rm light}} \frac{\hbar\omega_a}{2} .
\end{equation}
Then it is the problem to explain why $-E_{\rm eff}$ is so small; why is it much smaller than its natural value
given by the right-hand side of (\ref{Eeff})?  

Clearly, this problem is analogous to the cosmological constant problem in semiclassical gravity 
\cite{weinberg,nobb,sahni,carroll_cc,padm_cc}. 
Eq.~(\ref{einst2}) multiplied with $g$ 
\begin{equation}\label{einst2g}
 -{\cal H}_{\rm heavy}g = {\cal H}_{\rm light}g - Eg 
\end{equation}
is analogous to the $00$-component of the Einstein equation which, in appropriate units,
can be written as
\begin{equation}\label{einst2true}
 G_{\mu\nu}=T_{\mu\nu}+\Lambda g_{\mu\nu} ,
\end{equation}
where $G_{\mu\nu}$ is the Einstein tensor depending only on gravitational degrees, $T_{\mu\nu}$ is the 
energy-momentum tensor of matter, and $\Lambda$ is the cosmological constant. In this analogy, 
``heavy'' degrees are analogous to the gravitational degrees, ``light'' degrees are analogous to the matter degrees, 
and the constant $-E$ is analogous to the cosmological constant. In the semiclassical approximation one performs 
a quantization of matter while keeping gravity classical, so (\ref{einst2true}) is replaced with
 \begin{equation}\label{einst3true}
 G_{\mu\nu}=\langle\Psi|T_{\mu\nu}|\Psi\rangle +\Lambda g_{\mu\nu} ,
\end{equation}  
the $00$-component of which is analogous to (\ref{einst3}) multiplied with $g$
\begin{equation}\label{einst3g}
 -{\cal H}_{\rm heavy}g = \langle\psi| {\cal H}_{\rm light} |\psi\rangle g - Eg .
\end{equation}
In particular, in the matter ground state $|\Psi\rangle=|0\rangle$ one finds 
a very large quantum contribution analogous to (\ref{Hvac}), so there is a large discrepancy between 
the value of cosmological constant defined by the quantum ground state and the small value of cosmological constant 
found from cosmological observations \cite{weinberg,nobb,sahni,carroll_cc,padm_cc}.

Within our model, it is not difficult to understand where the problem comes from 
and how it should be resolved. In the diffeomorphism invariant action (\ref{A2}),
the constant energy $-E$ has physical consequences because it is coupled to $g$ via the term proportional to $\sqrt{g}E$.
This is analogous to the cosmological constant coupled to gravity via the term proportional to
$\sqrt{|\det{g_{\mu\nu}}|} \Lambda$. On the other hand, the action (\ref{A}) with (\ref{L}) is not diffeomorphism invariant 
and hence does not contain $\sqrt{g}$. As a consequence, adding a constant $E$ to the Lagrangian (\ref{L})
does not have any physical consequences. In the corresponding quantum theory described by the Schr\"odinger equation 
(\ref{sch}), the Hamiltonian is shifted by a constant value $-E$, which changes the phase of the quantum state 
by an additional phase factor $e^{iEt/\hbar}$, which does not have any physical consequences. 
The quantum ground state energy further shifts this value from $E$ to $E_{\rm eff}$ as given by (\ref{Eeff0}), 
but the new phase factor $e^{iE_{\rm eff}t/\hbar}$ still does not have any physical consequences.

Hence the conclusion is very similar to that in Sec.~\ref{SECtimeqg}.
The toy version of the cosmological constant problem originates from taking the 
diffeomorphism invariance too seriously. When one takes into account that this invariance is only emergent 
at the classical level, while fundamental quantum theory does not have this invariance, the toy cosmological
constant problem disappears in an obvious way.   

\section{Black hole and firewall} 
\label{SECbh}

\subsection{The model}

Consider a subsystem described by only two degrees of freedom $q(t)=\{ x(t),y(t) \}$, and suppose that the subsystem 
is invariant under rotations in the $x$-$y$ plane. Suppose also that $E=0$. Under these conditions, 
the action (\ref{A2}) reduces to
\begin{equation}\label{A2xy}
 \tilde{A}=\int dt\sqrt{g} \left[ \frac{m(\dot{x}^2 + \dot{y}^2)}{2g}-V(x,y) \right] ,   
\end{equation}
where $V(x,y)=V(x^2+y^2)$. Due to the rotational symmetry, it is convenient to work in polar coordinates
\begin{equation}
 z=\sqrt{x^2+y^2} , \;\;\;\; \varphi={\rm arctg}\frac{y}{x} ,
\end{equation}
with ranges
\begin{equation}\label{zpositive}
 z\in [0,\infty),  \;\;\;\; \varphi\in [0,2\pi) ,
\end{equation}
where the values $\varphi=0$ and $\varphi=2\pi$ are identified. Note that $z$ is the usual radial coordinate, but we denote it with 
$z$, rather than with $r$, for the reasons that will become clear later. Thus the action (\ref{A2xy}) can be written as
\begin{equation}\label{A2zvarphi}
 \tilde{A}=\int dt\sqrt{g} \left[ \frac{m(\dot{z}^2+z^2\dot{\varphi}^2)}{2g}-V(z^2) \right] ,   
\end{equation}
and the corresponding constraint (\ref{constr3}) reduces to
\begin{equation}\label{constr3z}
 \frac{m(\dot{z}^2+z^2\dot{\varphi}^2)}{2g} + V(z^2) =0 .
\end{equation}

To get an interesting solution of the constraint, let us suppose that the potential $V(z^2)$ for small $z$
has a form of an {\em inverted} harmonic oscillator
\begin{equation}\label{Vz}
 V(z^2)=-\frac{kz^2}{2} ,
\end{equation}
with $k>0$.
Thus, assuming in addition that $\varphi(t)=0$ and choosing the gauge
\begin{equation}\label{gauge}
 g(t)=1,
\end{equation}
the constraint (\ref{constr3z}) finally reduces to
\begin{equation}\label{constr3z2}
 \frac{m\dot{z}^2}{2} - \frac{kz^2}{2}=0 ,
\end{equation} 
which is a differential equation for $z(t)$
\begin{equation}\label{constr3z2d}
\left(\frac{dz(t)}{dt} \right)^2 = \gamma^2 z^2(t) ,
\end{equation}
where $\gamma=\sqrt{k/m}$. We will see that (\ref{constr3z2d}) 
describes a motion analogous to the radial motion of a particle around a black hole 
with a horizon at $z=0$. 

\subsection{Analogy with a black hole}

The solution of the differential equation (\ref{constr3z2d}) is
\begin{equation}\label{solution}
 z(t)=z(0)e^{\pm \gamma t} .
\end{equation}
The solution $z(t)=z(0)e^{-\gamma t}$ can be visualized as radial infalling towards $z=0$.
The infalling exponentially slows down as $z=0$ is approached, and it takes an infinite time $t$ to reach $z=0$.
Likewise, the solution $z(t)=z(0)e^{\gamma t}$ is a time inversion of the infalling, it describes an escaping 
from small $z$ towards $z\to\infty$. However, if it starts from $z(0)=0$, then it can never escape; 
it remains trapped at $z(t)=0$ forever. 
This behavior is very much analogous to infalling towards the black hole, or escaping from it. 
In particular, it takes an infinite time to reach the black hole horizon, from the point of view of observer staying  
at a fixed non-zero distance from the horizon. Also, 
an object initially at the horizon can never escape from it.
We see that the point $z=0$ is analogous to the black hole horizon.

Moreover, the analogy with black holes does not stop here. The solution (\ref{solution}) is obtained in the gauge 
(\ref{gauge}), but the theory is diffeomorphism invariant under time reparametrizations (\ref{t'}). 
Thus we can introduce a new time variable $t'$ defined implicitly by
\begin{equation}\label{tgamma'}
 e^{-\gamma t} = 1-\gamma t' ,
\end{equation}
so the infalling solution $z(t)=z(0)e^{-\gamma t}$ can be written as
\begin{equation}\label{zt'}
 z(t(t'))=z(0) [1-\gamma t']. 
\end{equation}
Now the point $z=0$ is reached after a {\em finite} time $t'=1/\gamma$. Furthermore, the solution 
(\ref{zt'}) can be extended to {\em negative} values of $z$ 
(this is the reason why we denote it with $z$, rather than with $r$), reached at times $t'>1/\gamma$. 
This is analogous to the Kruskal extension (see e.g. \cite{mtw,wald,carroll})
of the Schwarzschild solution in general relativity, where in appropriate spacetime coordinates 
a freely falling object reaches the horizon after a finite time and the Schwarzschild solution is extended 
{\em beyond} the horizon, thus describing not only the black hole exterior, but also its {\em interior}.  
Hence, the region of negative $z$ in the toy model is analogous to the black hole interior behind the 
Schwarzschild horizon.


\subsection{Effective spacetime}

The analogy above can also be made more explicit by introducing an effective spacetime metric.
The constraint (\ref{constr3z2d}) can be written as $\gamma^2 z^2 dt^2-dz^2=0$, 
which can be interpreted as motion of a relativistic massless particle in a spacetime with the effective metric
\begin{equation}\label{dseff}
ds^2_{\rm eff}=\Omega(t,z)[\gamma^2 z^2 dt^2-dz^2] ,
\end{equation}
where $\Omega(t,z)>0$ is an arbitrary conformal factor. This effective metric 
has a horizon at $z=0$. In particular, the metric in the square bracket
has the same form as the Rindler metric \cite{rindler,mtw}
\begin{equation}
ds^2_{\rm Rindler}=a^2 z^2 dt^2-dz^2 ,
\end{equation}
associated with an observer at $z=1/a$ accelerating with proper acceleration $a$.
The Rindler horizon at $z=0$ is known to have many similarities with the 
black hole horizon \cite{rindler,bd,mukhanov}. 

Since (\ref{dseff}) has a coordinate singularity at $z=0$, we want to see 
what happens with this singularity after the coordinate transformation (\ref{tgamma'}).
By applying (\ref{tgamma'}) to (\ref{dseff}), we get 
\begin{equation}\label{dseff'}
ds^2_{\rm eff}=\Omega \left[\frac{\gamma^2 z^2 dt'^2}{(1-\gamma t')^2}-dz^2\right] ,
\end{equation}
which is still singular at $z=0$. However, the singular quantity
\begin{equation}
g'_{00}=\frac{\Omega\gamma^2 z^2}{(1-\gamma t')^2} 
\end{equation}
is in fact regular along the infalling trajectory (\ref{zt'}), i.e.
\begin{equation}\label{dseff'2}
g'_{00} \stackrel{\rm traj}{=}\Omega\gamma^2 z^2(0)
\end{equation} 
is regular provided that the initial position obeys $z(0)\neq 0$.

A standard way to completely remove the coordinate singularity at the horizon $z=0$
is to introduce the new spacetime coordinates 
\begin{equation}\label{mink}
T=z\, {\rm sh} \gamma t, \;\;\; Z=z\, {\rm ch} \gamma t. 
\end{equation}
Indeed, an elementary calculus shows that
$dT^2-dZ^2=\gamma^2 z^2 dt^2-dz^2$, so (\ref{dseff}) can be written as 
\begin{equation}\label{dseffMink}
 ds^2_{\rm eff}=\Omega[dT^2-dZ^2] .
\end{equation}
In these coordinates the relativistic massless particle obeys $dT^2-dZ^2=0$, so the infalling solution is 
\begin{equation}\label{ZT}
 Z(T)=Z(0)-T ,
\end{equation}
which corresponds to (\ref{zt'}). 

Now we want to express the position of the horizon $z=0$ in the $T,Z$ coordinates.
Inserting $z=0$ into (\ref{mink}) gives $(T,Z)=(0,0)$, if $t$ is finite. But what about the limit $t\to\pm\infty$?
In this limit (\ref{mink}) gives $Z/T=\pm 1$ for any $z$, including the limit $z\to 0$, so the two lines
$Z=\pm T$ are also consistent with $z=0$. Thus the horizon is the union of the point $(T,Z)=(0,0)$ 
(corresponding to finite $t$) and the lines $Z=\pm T$ (corresponding to $t\to\pm\infty$). 
But this union is simply the two lines $Z=\pm T$, so we conclude that the horizon {\em is} the two lines $Z=\pm T$. 
The line $Z=T$ is the future horizon, which is characteristic for a black hole,
while the line $Z=-T$ is the past horizon, which is characteristic for a white hole.

Thus we see that the infalling solution (\ref{ZT}) 
crosses the future horizon $Z=T$ and extends beyond the future horizon, which corresponds 
to the extension beyond the analogue horizon $z=0$ in (\ref{zt'}).

Finally note that the effective spacetime metric can be introduced not only for the potential (\ref{Vz}), 
but also for any potential $V(x,y)$ in (\ref{A2xy}), provided that it is negative. The constraint resulting
from (\ref{A2xy}) is
\begin{equation}
 \frac{m(\dot{x}^2 + \dot{y}^2)}{2g}+V(x,y)=0 ,
\end{equation}
which in the gauge $g=1$ can be written as
\begin{equation}
 -\frac{2V(x,y)}{m}dt^2-dx^2-dy^2=0 .
\end{equation}
This can be interpreted as motion of a relativistic massless particle in a spacetime with the effective metric
\begin{equation}
 ds^2_{\rm eff}=\Omega(t,x,y) \left[ -\frac{2V(x,y)}{m}dt^2-dx^2-dy^2 \right] ,
\end{equation}
where $\Omega(t,x,y)>0$ is an arbitrary conformal factor.
This metric has the relativistic signature  $(+--)$, provided that $V(x,y)<0$. 
Taking $\Omega=1$ for convenience and 
defining the effective ``Newtonian'' gravitational potential $\phi_{\rm grav}(x,y)$ 
through the standard relation \cite{carroll}
\begin{equation}
 g_{00}(x,y)=1+2\phi_{\rm grav}(x,y) ,
\end{equation}
we see that the potentials $V$ and $\phi_{\rm grav}$ are related as 
\begin{equation}\label{phigrav}
 \phi_{\rm grav}(x,y)=-\frac{V(x,y)}{m} -\frac{1}{2} .
\end{equation}
The important message of (\ref{phigrav}) is that $\phi_{\rm grav}$ corresponds to $-V$, 
rather than to $V$ as one might naively expect. In particular,  
we see that a {\em repulsive} potential $V$ such as (\ref{Vz}) corresponds to an 
{\em attractive} gravitational potential $\phi_{\rm grav}$.

\subsection{The firewall}
\label{SECfirewall}

We have seen that the solution (\ref{zt'}) can be extended to negative values of $z$,
and that this extension is analogous to the extension of black hole behind the horizon.
However, in the toy model, the extension is conceptually problematic.
How can the extension to negative values of $z$ be compatible with the fact 
that the $z$-coordinate was restricted to non-negative values by {\em definition}, in Eq.~(\ref{zpositive})?
The answer is that it cannot! Only non-negative values of $z$ are physical.
The region of space with negative $z$ does not exist. The motivation for extension to negative values of $z$
has arisen from (\ref{zt'}), which, in turn, has arisen from a new time coordinate introduced in (\ref{tgamma'}). 
But the original model (\ref{A}) with (\ref{L}) is not diffeomorphism invariant, i.e. it does not allow 
arbitrary redefinitions of the time coordinate. From this point of view, the gauge (\ref{gauge}) is not merely an arbitrary 
choice, but the correct physical value of $g$.
The negative values of $z$ have arisen from taking the diffeomorphism invariance too seriously, 
while this invariance is just an emergent feature resulting from a formalism
that encoded the classical value of energy $E$ into the action, as described in Sec.~\ref{SECemerg}.
 
The conclusion above that there is no region behind $z=0$ is completely classical,
it does not involve any quantum physics. Nevertheless, a semiclassical version resembling 
Hawking radiation can also be constructed. 
Suppose that two entangled particles are created at $z>0$, one infalling and the other 
escaping, thus mimicking the Hawking pair. Suppose also that the potential $V(z^2)$, given by (\ref{Vz}) for small $z$, 
is defined for all $z\geq 0$ as
\begin{equation}\label{Vzall}
 V(z^2)=\left\{ 
\begin{array}{ll}
 -kz^2/2 &  {\rm for} \;\; z\leq z_0  \\
 -V_0 & {\rm for} \;\;  z\geq z_0 ,
\end{array}
\right.  
\end{equation}
where
\begin{equation}\label{z0}
 z_0=\sqrt{\frac{2V_0}{k}} .
\end{equation}
This potential can be visualized as a flat valley at the constant potential $-V_0$ for $z>z_0$, with a 
hill of height $V_0$, radius $z_0$, and the top at $z=0$. It mimics a stationary black hole approximated 
with flat geometry for $r\geq r_0$, which is justified if $r_0$ is much larger than the Schwarzschild radius.
To mimic a non-stationary evaporating black hole, we modify (\ref{Vzall}) and (\ref{z0}) to
 \begin{eqnarray}\label{Vzallt}
V(z^2,t) &=& \left\{ 
\begin{array}{ll}
 -k(t)z^2/2 & {\rm for} \;\; z< z_0(t)  \\
 -V_0 & {\rm for} \;\;  z\geq z_0(t) ,
\end{array}
\right.  
\\
z_0(t) &=& \sqrt{\frac{2V_0}{k(t)}} , 
\end{eqnarray}   
where $k(t)$ is an increasing function that, after a large but finite time $t_*$, becomes infinite $k(t_*)=\infty$. 
Thus the radius $z_0(t)$ shrinks and becomes zero at time $t_*$, which mimics the shrinking of the 
evaporating black hole. The information paradox can now be formulated as follows.
The peak of the infalling wave packet follows approximately the classical trajectory (\ref{zt'}), thus entering the 
region behind $z=0$, i.e. behind the top of the hill. 
But at late times $t>t_*$ the potential is $V(z^2)=-V_0$, so there is no hill and hence no region behind the 
top of the hill. It looks as if the infalling particle disappears at late times, so the 
remaining escaping particle in the mixed state
seems to contradict unitarity of quantum mechanics. This is the toy version of the black hole information paradox.
The solution of the paradox is that the region behind $z=0$ never existed in the first place. As we said, 
the motivation for extension to negative values of $z$
originated from (\ref{zt'}), which, in turn, originated from introducing a new time coordinate in (\ref{tgamma'}),
which, however, is not allowed in the fundamental theory without diffeomorphism invariance.
 
Remarkably, the non-existence of the region behind $z=0$ in the toy model has an analogy 
in black hole physics. With a motivation to resolve the black hole information paradox 
\cite{gid,har,pres,pag,gid2,str,math1,math2,hoss,dundar,harlow,polchinski,chakra,marolf,fabbri}
in semiclassical gravity, it has been proposed that the black hole interior does not exist; 
the black hole horizon represents a physical boundary called {\em firewall} \cite{AMPS,apologia,dundar}. 
The problem with the firewall is to reconcile it with standard classical general relativity, which  
predicts that the black hole interior exists, and that the horizon is not a physical boundary.   
But such a standard view of classical general relativity is a consequence of the 4-dimensional diffeomorphism invariance. 
Alternatively, if the
4-dimensional diffeomorphism invariance in general relativity is emergent in a way similar to the 
emergence of the 1-dimensional diffeomorphism invariance in our toy model, then the 4-dimensional diffeomorphism invariance
should not be taken too seriously even in the classical theory. If so, then the existence of the 
black hole interior resulting from the Kruskal extension should not be trusted. Such an alternative view of classical gravity,
if correct, makes the firewall perfectly compatible with classical physics, which resolves the firewall problem.

Hence the conclusion is similar to that in Secs.~\ref{SECtimeqg} and \ref{SECcc}.
The toy version of the firewall problem originates from taking the 
diffeomorphism invariance too seriously. When one takes into account that this invariance is only emergent, 
while the fundamental theory does not have this invariance, the toy firewall problem
disappears in an obvious way. 

\section{Towards emergent 4-dimensional diffeomorphism invariance}
\label{SEC4dim}

The motivation for studying the toy models with 1-dimensional diffeomorphism invariance is to teach us something 
about the real 4-dimensional diffeomorphism invariance, namely, about real classical, semiclassical and quantum gravity.
So the question is, how the ideas of the toy models can be generalized to 4-dimensional diffeomorphism invariance?
Unfortunately, we do not have a full answer to that question. A full answer would be tantamount to having a full theory 
of quantum gravity, which, of course, we do not have. 
Nevertheless, inspired by the toy models, we  
sketch an idea how such a generalization {\em might} look like. What we present here can be thought of as a gist of a
research program based on a series of educated guesses\footnote{``Educated guess'' is (supposed to be) 
a well balanced term, between the over-pretentious ``conjecture'' and over-cynical ``wishful thinking''.}, 
which at the current level is very far from a fully developed theory. 

Our starting point of view is that the spacetime curvature emerges from a massless spin-2 field
\cite{feynman_lecture,weinberg_spin2,thirring,nik_nongeom,deser10}, and not the other way around.
Roughly, this means that in the formula
\begin{equation}\label{getaphi}
g_{\mu\nu}(x)=\eta_{\mu\nu}+\phi^{\text{spin-2}}_{\mu\nu}(x) ,
\end{equation}
relating the curved spacetime metric $g_{\mu\nu}(x)$ to the flat Minkowski metric $\eta_{\mu\nu}$
and the massless spin-2 field $\phi^{\text{spin-2}}_{\mu\nu}(x)$, the quantities on the right-hand 
side are more fundamental than that on the left-hand side.
Philosophically, such a view complies much better with string theory than with loop quantum gravity.
In the fundamental theory, the formula (\ref{getaphi}) is expected to be valid only in some approximative sense.

We assume that there is some fundamental action $A[\phi]$ without 
diffeomorphism invariance, where $\phi=\phi(x)$ is a collective symbol for all fundamental dynamical fields
\begin{equation}
 \phi=\{\phi_{\text{matt}},\phi_{\text{spin-2}},\ldots\} .
\end{equation}
Here $\phi_{\text{matt}}$ are the usual ``matter'' fields of spins 0, $\frac{1}{2}$ and 1, the field $\phi_{\text{spin-2}}$ 
is the massless spin-2 field, and the ellipses are possible other fields beyond the Standard Model of particle physics. 
The $x$ denotes a spacetime position in 4 or more dimensions. From the action $A[\phi]$ one can derive
the symmetrized energy-momentum tensor $T_{\mu\nu}[\phi;x]$, 
which is conserved when the equations of motion 
\begin{equation}\label{eom}
\delta A/\delta\phi(x)=0
\end{equation} 
are satisfied. In classical physics 
the fields $\phi(x)$ attain some definite values $\Phi(x)$, where $\Phi(x)$
is a definite solution of (\ref{eom}). Thus we can define 
\begin{equation}
E_{\mu\nu}(x)\equiv T_{\mu\nu}[\Phi;x] , 
\end{equation}
which is a generalization of the definite energy $E$ appearing in (\ref{constr2}).
For example, in a classical vacuum in Minkowski spacetime, 
the $E_{\mu\nu}(x)$ may take the form 
\begin{equation}
E_{\mu\nu}(x)=-\Lambda\eta_{\mu\nu} ,
 \end{equation}
where $\Lambda$ is a constant. 
But whatever the $E_{\mu\nu}(x)$ is, in classical physics
we can always write
\begin{equation}\label{Tmunu}
 T_{\mu\nu}[\phi;x]-E_{\mu\nu}(x)=0 ,
\end{equation}
which is a generalization of (\ref{constr2}). 
In some limit one expects that $T_{\mu\nu}[\phi;x]$ can be decomposed as
\begin{equation}
T_{\mu\nu}[\phi;x]=T^{\text{ matt}}_{\mu\nu}[\phi;x] + T^{\text{ spin-2}}_{\mu\nu}[\phi;x] + \ldots .
\end{equation}
With this decomposition, 
(\ref{Tmunu}) looks very much like the Einstein equation (\ref{einst2true}) written in 
the non-geometric spin-2 language.

Now the idea is to think of (\ref{Tmunu}) as a constraint derived from a new action $\tilde{A}[\phi,g]$, 
where $g(x)=\{ g_{\mu\nu}(x) \}$ is a symmetric tensor field. By analogy with (\ref{A2}),
one expects that the new action $\tilde{A}[\phi,g]$ is diffeomorphism invariant, so that 
the diffeomorphism-covariant equation 
\begin{equation}
\delta\tilde{A}/\delta g_{\mu\nu}(x)=0
\end{equation}
reduces to (\ref{Tmunu}) when the gauge for $g_{\mu\nu}$ is chosen appropriately.
One also expects that, in a certain limit, the action $\tilde{A}[\phi,g]$
reduces to the usual gravitational action with the matter term, 
the Einstein-Hilbert term, and the cosmological term.
This is, roughly, how the 4-dimensional 
diffeomorphism invariance is expected to emerge at the classical level. However, 
the fundamental action that needs to be quantized in this scheme is $A[\phi]$, not $\tilde{A}[\phi,g]$.

With this approach, it it easy to see that there is no problem of time in quantum gravity, simply because 
the fundamental action $A[\phi]$ does not have a Hamiltonian constraint. The Hamiltonian $H$ derived from 
$A[\phi]$ does not need to vanish on-shell. Likewise, there is no cosmological constant problem, in the sense that energy 
(associated with $H$) of the quantum ground state does not have physical consequences.
Finally, the quantum time evolution defined by $e^{-iHt/\hbar}$ is unitary, so all quantum processes, including 
Hawking radiation, are compatible with unitarity. Nevertheless, at this level, it is not clear how exactly 
the information paradox associated with Hawking radiation resolves. Since the quantum theory lacks 
diffeomorphism invariance, the firewall scenario discussed in Sec.~\ref{SECfirewall} scenario seems plausible. 
In the same spirit, since quantum gravity is not fundamentally geometrical in this picture, 
inherently geometrical proposals involving wormholes, such as ER=EPR \cite{er=epr} and black hole islands \cite{island}, 
seem less plausible. Nevertheless, at the current level of understanding of the ideas sketched above,
it is impossible to make definite precise claims about the quantum nature of black holes. 
 
\section{Discussion and conclusion}
\label{SECconcl}

In this paper we have constructed 
toy versions of the problem of time in quantum gravity, of the cosmological constant problem,
and of the black hole firewall problem. Within the models, 
the problems originate from taking the 1-dimensional diffeomorphism invariance 
too seriously. This 1-dimensional diffeomorphism invariance, realized as time-reparametrization invariance,
is emergent, rather than fundamental, 
and when one takes it into account the problems disappear in a rather natural way. 
The problem of time disappears because quantum energy is uncertain in the absence of fundamental time-reparametrization invariance. 
The cosmological constant problem disappears because a shift of energy by a constant does not have physical 
consequences in the absence of fundamental time-reparametrization invariance.
The black hole firewall problem disappears because a firewall at the horizon 
may be completely compatible with classical physics when the diffeomorphism invariance is interpreted as emergent, 
rather than fundamental.

We stress that our resolution of the problem of time in Sec.~\ref{SECtimeqg}
requires that the whole Universe is in a state of uncertain energy. 
Can this requirement be relaxed? Let us discuss various possibilities, 
together with their shortcomings. 
Naively one might think that the whole Universe should have a well defined energy,
but that a subsystem $S$ can still have a non-trivial time 
dependence because the subsystem $S$ does not need to have a well defined energy.
However, if the whole Universe is time-independent, then any subsystem of it is also time-independent. 
To see this explicitly, suppose that the whole Universe is in the energy eigenstate
$|\psi(t)\rangle =e^{-iEt/\hbar}|\psi(0)\rangle$. Then the density matrix of the whole 
Universe is
\begin{equation}
\rho(t)=|\psi(t)\rangle \langle\psi(t)|=|\psi(0)\rangle \langle\psi(0)|=\rho(0) ,
\end{equation} 
which is clearly time-independent. 
Hence the state $\rho_S$ of the subsystem $S$ is given by the partial trace
over the rest $R$ of the Universe (defined by all degrees of freedom except those of $S$)
\begin{equation}\label{part_trace}
\rho_S={\rm Tr}_R\rho(0) .
\end{equation} 
Clearly, the right-hand side of (\ref{part_trace}) is time-independent, for any 
decomposition of the whole Universe into a subsystem $S$ and the rest $R$.
Indeed, if the energy of the whole Universe is well defined, then
the subsystem $S$ can have uncertain energy only when it is an open system 
entangled with $R$, but the state of the open system is not described by the 
Schr\"odinger equation, so the uncertainty of its energy does not imply time dependence.
Open systems often behave classically due to decoherence caused by the environment  
\cite{schloss}, but decoherence itself is a time-dependent process, 
so there cannot be any decoherence if the whole Universe is time-independent. 
All this shows that a subsystem cannot depend on the time $t$ if the whole Universe 
does not depend on $t$. When the wave function of the Universe is an energy eigenstate,
a $t$-dependence can be incorporated by various non-minimal interpretations
of quantum mechanics \cite{nik_miniatures}, e.g. by assuming that macroscopic 
classicality is fundamental (rather than emergent from quantum mechanics),
or by postulating additional $t$-dependent variables, or 
by modifying the Schr\"odinger equation, but any such quantum interpretation 
introduces an additional level of controversy. Finally,
it is possible to associate a dependence on ``time"
with a subsystem by redefining the notion of ``time" itself. The best known example 
is the Page-Wootters ``time" \cite{page} (with many variations, such as 
\cite{timeobs3,timeobs4}), based on the idea that a wave function which does not 
depend on the external time $t$ may still depend on the configuration variable 
$q_c$ corresponding to the clock observable, suggesting that 
$q_c$ itself can be interpreted as ``time". Such approaches are interesting 
in its own right, but are orthogonal to the approach of the present paper.   
Let us just say that, in our view, it is not clear why would a dependence 
on $q_c$ be interpreted as a time {\em evolution}, in a sense in which we usually associate evolution 
with the dependence on $t$. To illustrate the problem, 
consider a $t$-independent wave function $\psi(q_c,q_p)$,
where $q_c$ is the position of a clock needle, 
while $q_p$ is the position of something else, say a pencil. 
Since the dependence of $\psi$ on $q_p$ is usually not interpreted as any kind of evolution, 
it is not clear why would the dependence on $q_c$ be interpreted so.
Hence we conclude that having the whole Universe in a state of uncertain energy is the most 
straightforward approach to explain the time evolution, while all other possibilities 
lead to additional problems.

Next note that the physical irrelevance of vacuum energy in the context of the cosmological constant problem
is compatible with the Casimir effect. 
The description of Casimir effect in terms of vacuum energy is just an effective macroscopic description, 
while the fundamental microscopic origin of Casimir effect lies in van der Waals forces \cite{jaffe,nik_casimir,nik_casimir_toy}.
In particular, it can be understood in terms of a toy model \cite{nik_casimir_toy} similar to that of the present paper.  

In our toy models, the solutions of the problems of time and of the cosmological constant are rather generic; 
the solutions do not depend on details of the models. In particular, even though the cosmological constant 
problem is discussed for quantum harmonic oscillators, the solution of the problem works in essentially the same way for 
any other interaction $V(q)$ that leads to a non-zero quantum ground state energy. 

By contrast, our solution 
of the toy black hole firewall problem is not so generic, it depends on details of the model.
Perhaps different models could suggest totally different solutions of the black hole information paradox, 
without any hints for the existence of firewalls. Or perhaps some models would describe 
classical states resembling black holes, but without any hints how to solve the information paradox.
More research is needed to better understand how the lack of fundamental diffeomorphism invariance
may, or may not, help to solve the information paradox.

More importantly, it is not at all clear whether such toy 1-dimensional  
ideas can, and should, be generalized to the real 4-dimensional diffeomorphism invariance of general relativity.
In Sec.~\ref{SEC4dim} we have sketched how such a generalization might look like,
but it is far from a fully developed theory. 
Nevertheless, the conceptual simplicity of solutions of the toy problems
seems suggestive, so we believe that this conceptual simplicity could at least serve as a source of inspiration for further research.

In any case, we believe that our analysis of the toy models with emergent diffeomorphism invariance may influence 
how physicists think about general relativity at an intuitive level. A change of intuition may also induce 
new technical results and, hopefully, contribute to better understanding of semiclassical and quantum gravity.     
     
\section*{Acknowledgements}
The author is grateful to T. Juri\'c for discussions.
This work was supported by the Ministry of Science of the Republic of Croatia.

\end{document}